# Mechanism of Information Transmission from a Spot Rate Market to Crypto-asset Markets

Takeshi YOSHIHRA[1], and Taisei KAIZOJI[2*]

[1]*DTrust Operations Department, State Street Trust and Banking Co., Ltd., Fukuoka 812-0036, Japan*
[2]*NGraduate School of Arts and Sciences, International Christian University, Mitaka 181-8585, Japan*

*E-mail: kaizoji@icu.ac.jp

We applied the SVAR-LiNGAM to illustrate the causal relationships among the spot exchange rate, and three crypto-asset exchange rates, Bitcoin, Ethereum, and Ripple. It was notable that the causal order, the EUR/USD spot rate→Bitcoin→Ethereum→Ripple, was obtained by this approach. All the instantaneous effects were strongly positive. Moreover, it was notable that Bitcoin can influence the EUR/USD spot rate positively with a one-day time lag.
**KEYWORDS:** the causal order, the crypto-asset exchange rates, the SVAR-LiNGAM

## 1. Introduction

Crypto-asset prices and their volatility have been increasingly drawing attention in these few years. Several analyses on Bitcoin and its price can be found in the recent econometric study. On the one hand, Cheah and Fry (2015) found a speculative bubble in the time-series price data of Bitcoin. Corbet, Lucey, and Yarovaya (2018) found that there were periods of bubble behavior in the historical price data of Bitcoin and Ethereum. On the other hand, Urquhart (2016) suggested that the Bitcoin markets were not efficient, but they were getting more efficient. Further, Nadarajah and Chu (2017) indicated that a power transformation model of Bitcoin returns seemed to be weakly efficient. Tiwari, Jana, Das, and Roubaud (2018) analyzed Bitcoin price data by using long-range dependence estimators, and reported that the market was efficient. Subsequently, Nan and Kaizoji (2019) compared Bitcoin with $EUR/USD$ spot, future, and forward rates, and concluded that weak and semi-strong efficiency of Bitcoin could hold in the long term.

Following these empirical investigations, we analyze the efficiency and causal relationships of Bitcoin, Ethereum, and Ripple by applying the Nan and Kaizoji's approach that compares exchange rates in the real markets to indirect crypto-asset exchange rates ($CER$). Its equation is given by

$$CER = \frac{CRP\_EUR}{CRP\_USD}$$

where $CRP\_EUR$ and $CRP\_USD$ are the crypto-asset prices in Euros and U.S. Dollars, respectively. The aim of this method is not only to eliminate the influence of the exponential growth of crypto-asset markets, but also to compare those markets with FX markets (see Figure 1 and 2). In the following chapters, we consider four time series variables, $SPT_t$ (the spot rate of $EUR/USD$), $BTC_t$ (the crypto-asset exchange rate by Bitcoin), $ETH_t$ (that by Ethereum), and $XRP_t$ (that by Ripple).

The rest of this paper is structured as follows. We present the theoretical backgrounds of this research in the next section. Subsequently, we show the results of empirical study and describe some implications for our findings. Conclusions are provided in the last section.

## 2. Data and methodology

*2.1 Data and statistical software*

Crypto-asset data, $BTC\_USD$, $BTC\_EUR$, $ETH\_USD$, $ETH\_EUR$, $XRP\_USD$, and $XRP\_EUR$, in this paper were closing price[1] from Yahoo! Finance. The $EUR/USD$ spot rate data was closing price from the Bank of England, and its missing values on weekends and annual holidays were complemented by the closing prices on Fridays. To eliminate the influence of arbitrage opportunity, we use crypto-asset and FX data in the same time zone. The original and complemented data of $SPT_t$ showed close numbers in summary statistics, and it is indicative of unbiasedness by this complementation (see Table 1). The following tests including $SPT_t$ were analyzed by the complemented data. The sample size of $Original\ SPT_t$ was 735, and that of $Weekend\text{-}filled\ SPT_t$, $BTC_t$, $ETH_t$, and $XRP_t$ was 1062. The sample period of the four processes was from December 04, 2016 to October 31, 2019. All statistical analyses proceeded with R (R Core Team, 2019). The original programming code for the VAR-LiNGAM estimation was given in Moneta, Entner, Hoyer, and Coad (2013), and Entner's website.

*2.2 Causal discovery and structural modeling*

*2.2.1 Limitations of VAR and VECM without instantaneous effects*

Generally, VAR and VECM approaches have been used to estimate the relationships among the variables. They are very convenient and widely used to predict the future prices in economics and finance. However, such conventional approaches might result in incorrect estimation or analysis due to neglecting instantaneous effects among variables (Hyvärinen, Zhang, Shimizu, & Hoyer, 2010). Consider the VAR (*p*) model of equation (1),

$$\boldsymbol{y}_t = \boldsymbol{\gamma} + \boldsymbol{\Pi}_1 \boldsymbol{y}_{t-1} + \cdots + \boldsymbol{\Pi}_p \boldsymbol{y}_{t-p} + \boldsymbol{\varepsilon}_t, \tag{1}$$

this is called as a "reduced form" of VAR. Alternatively, we also can think of a VAR model which includes instantaneous effects by adding a coefficient $\boldsymbol{B}_0$ for $\boldsymbol{y}_t$,

$$\boldsymbol{B}_0 \boldsymbol{y}_t = \boldsymbol{c} + \boldsymbol{B}_1 \boldsymbol{y}_{t-1} + \boldsymbol{B}_2 \boldsymbol{y}_{t-2} + \cdots + \boldsymbol{B}_p \boldsymbol{y}_{t-p} + \boldsymbol{u}_t, \tag{2}$$

where $\boldsymbol{B}$ is a matrix of coefficients for $\{\boldsymbol{y}_t, \boldsymbol{y}_{t-1}, \boldsymbol{y}_{t-2}, \cdots, \boldsymbol{y}_{t-p}\}$. Equation (2) is known as a "structural form" of VAR, or SVAR.

Hyvärinen et al. (2010) noted that there were some estimation errors if we did not consider an instantaneous effect $\boldsymbol{B}_0$. For example, lagged effects estimated in the conventional VAR are actually instantaneous. Consider a bivariate SVAR (1) model,

$$\begin{pmatrix} 0 & 1 \\ 0 & 0 \end{pmatrix} \boldsymbol{y}_t = \begin{pmatrix} 0.9 & 0 \\ 0 & 0.9 \end{pmatrix} \boldsymbol{y}_{t-1} + \boldsymbol{u}_t. \tag{3}$$

This equation indicates that $y_{2t}$ has an instantaneous effect to $y_{1t}$, but does not have such a lagged effect. On the other hand, if we model the same data by a VAR (1), we obtain

$$\boldsymbol{y}_t = \begin{pmatrix} 0.9 & 0.9 \\ 0 & 0.9 \end{pmatrix} \boldsymbol{y}_{t-1} + \boldsymbol{\varepsilon}_t. \tag{4}$$

As this VAR model neglects the instantaneous effect, the estimated coefficients are all lagged. This could lead to a crucial misunderstanding when we make an inference from the results obtained.

Hyvärinen et al. (2010) provided another example for the problematic interpretation of a VAR model. Consider a three-variable SVAR (1) model

$$\begin{pmatrix} 0 & 0 & 0 \\ 1 & 0 & 0 \\ 0 & 1 & 0 \end{pmatrix} \boldsymbol{y}_t = \begin{pmatrix} 0.9 & 0 & 0 \\ 0 & 0.9 & 0 \\ 0 & 0 & 0.9 \end{pmatrix} \boldsymbol{y}_{t-1} + \boldsymbol{u}_t. \tag{5}$$

This equation represents that $y_1$ and $y_2$ have instantaneous effects on $y_2$ and $y_3$, respectively, and that no causal relation, except for autocorrelating coefficients $y_{i,t-1} \to y_{it}$ of each variable, exist in the lagged coefficient matrix. Here, if we construct a VAR (1) model for the same data, the equation is given by

$$\begin{aligned} \boldsymbol{y}_t &= \left\{ \begin{pmatrix} 1 & 0 & 0 \\ 0 & 1 & 0 \\ 0 & 0 & 1 \end{pmatrix} - \begin{pmatrix} 0 & 0 & 0 \\ 1 & 0 & 0 \\ 0 & 1 & 0 \end{pmatrix} \right\}^{-1} \begin{pmatrix} 0.9 & 0 & 0 \\ 0 & 0.9 & 0 \\ 0 & 0 & 0.9 \end{pmatrix} \boldsymbol{y}_{t-1} + \boldsymbol{\varepsilon}_t \\ &= \begin{pmatrix} 0.9 & 0 & 0 \\ 0.9 & 0.9 & 0 \\ 0.9 & 0.9 & 0.9 \end{pmatrix} \boldsymbol{y}_{t-1} + \boldsymbol{\varepsilon}_t. \end{aligned} \tag{6}$$

If making an inference on the coefficients in this VAR, we would indicate that $y_1$ has a direct effect to $y_3$. However, this is spurious because direct causality does not exist in the original SVAR model. Here, the conventional VAR (and VECM) approach sometimes leads to incorrect analyses.

*2.2.2 SVAR model*

The structural forms have one crucial obstacle when it comes to their estimation. When estimating the simultaneous equations by OLS, we obtain a result with a bias since the dependent variables and error terms are correlated. Therefore, the reduced-form VAR is estimated at the beginning. Considering the SVAR model of equation (2), when multiplying the both sides of equation (2) by $\boldsymbol{B}_0^{-1}$, we obtain

$$\begin{aligned} \boldsymbol{y}_t &= \boldsymbol{B}_0^{-1}\boldsymbol{c} + \boldsymbol{B}_0^{-1}\boldsymbol{B}_1\boldsymbol{y}_{t-1} + \boldsymbol{B}_0^{-1}\boldsymbol{B}_2\boldsymbol{y}_{t-2} + \cdots + \boldsymbol{B}_0^{-1}\boldsymbol{B}_p\boldsymbol{y}_{t-p} + \boldsymbol{B}_0^{-1}\boldsymbol{u}_t \\ &= \boldsymbol{\gamma} + \boldsymbol{\Pi}_1\boldsymbol{y}_{t-1} + \cdots + \boldsymbol{\Pi}_p\boldsymbol{y}_{t-p} + \boldsymbol{\varepsilon}_t, \end{aligned} \tag{7}$$

where $\boldsymbol{\gamma} = \boldsymbol{B}_0^{-1}\boldsymbol{c}$, $\boldsymbol{\Pi}_t = \boldsymbol{B}_0^{-1}\boldsymbol{B}_t$, and $\boldsymbol{\varepsilon}_t = \boldsymbol{B}_0^{-1}\boldsymbol{u}_t$. It should be now noted that equation (7) is identical to equation (2), which is the conventional reduced-form VAR model. Equation (7) can be estimated by OLS without a bias. Second, we should consider converting from the reduced form to the structural form. However, it is not necessarily possible to identify the structural form from the reduced form because the structural form has $n(n-1)/2$ parameters more than the reduced form. This is known as an identification problem of the structural form.

The identification problem is generally solved by assuming a causal order of variables, or by restricting parameters based on economics or finance theory. The former assumed model is called a recursive SVAR, and the latter is known as a non-recursive SVAR. For the recursive model, one of the typical ways to iron out the identification problem is to orthogonalize[3] the structural-form error term $\boldsymbol{u}_t$ (Kilian & Lütkepohl, 2017). In other words, $\boldsymbol{B}_0$ is assumed to be a lower triangular matrix whose diagonal components are equal to 1. For the non-recursive one, there are several types of restrictions that can identify the model. For instance, Blanchard and Quah (1989) imposed long-term restrictions on the structural forms. Identification by long-term restrictions allows us to identify whether the processes have structural shocks; however, there are a number of limitations for this methodology, such as robustness for low-frequency data, and data transformation sensitivity of whether the variables are computed in level or differences (Kilian & Lütkepohl). The recursive model and the non-recursive model with long-term restrictions are the practical ways to analyze. However, they need to assume causal order among the variables, else they will face inevitable constraints imposed by the volume of data processing. As the main purpose of this paper is to clarify causal relationships without a priori assumptions, a more data-driven method is needed to facilitate the analysis of causality.

On the other hand, Bernanke (1986) and Sims (1986) applied short-term restrictions to the structural forms. However, Kilian and Lütkepohl (2017) noted that identification by short-term restrictions is challenging, because it is sometimes difficult to impose restrictions which completely identify the SVAR. Swanson and Granger (1997) proposed one way to clear this up by combining structural analysis with graph that represents causal relationships using a network structure. Subsequently, Moneta, Entner, Hoyer, and Coad (2013) proposed a graph-theoretic structural analysis using a data-driven method called "independent component analysis (ICA)." In the following sections, ICA and LiNGAM methods are introduced in order to give a theoretical explanation of our application of an SVAR-LiNGAM approach.

*2.2.3 Independent component analysis and non-Gaussianity*

According to Hyvärinen, Karhunen, and Oja (2001), ICA can be defined as a methodology to uncover unknown elements from multivariate continuous-value data. One of the unique points of this analytical method is to focus on the elements that are non-Gaussian and statistically independent from other elements. Consider an example of a structural equation model provided by Shimizu (2017),

$$y_1 = a_{11}s_1 + a_{12}s_2$$
$$y_2 = a_{21}s_1 + a_{22}s_2$$

where $s$ denotes an unobserved independent component, and $a$ denotes an unobserved coefficient which mixes $s$. This structural model can be written in a matrix representation as

$$\begin{bmatrix} y_1 \\ y_2 \end{bmatrix} = \begin{bmatrix} a_{11} & a_{12} \\ a_{21} & a_{22} \end{bmatrix} \begin{bmatrix} s_1 \\ s_2 \end{bmatrix}. \tag{8}$$

Figure 3 shows a causal graph of this structural model. For general cases, the *p*-variable ICA model for $y_i$ can be defined as

$$y_i = \sum_{j=1}^{n} a_{ij} s_j \quad (i,j = 1,2,\cdots,n) \tag{9}$$

where $a_{ij}$ is the mixing coefficient, and $s_j$ is defined as statistically mutually independent. ICA estimates both the coefficient $a_{ij}$ and the independent component $s_j$ when $y_i$ is observed.

Hyvärinen et al. (2001) imposed two restrictions to make estimation possible. The first was the assumption that $s_j$ was statistically independent. In other words, this assumption can be represented as

$$p(s_1, s_2, \cdots, s_n) = p_1(s_1)p_2(s_2)\cdots p_n(s_n)$$

where $p(\cdot)$ denotes a probability density function. This assumption is not too strict because the effect of dependent elements is considered in the coefficient term $a_{ij}$. The second was that $s_j$ followed non-Gaussian distributions. Hyvärinen et al. noted that Gaussian distributions did not have much essential information to estimate the model. Consider a probabilistic variable $X$ that follows a Gaussian distribution, its probability density function $p(X)$ can be obtained as

$$p(X) = \frac{1}{\sqrt{2\pi}\sigma} \exp\left(-\frac{(X-\mu)^2}{2\sigma^2}\right).$$

Subsequently, the moment generating function $M(t)$ can be represented by

$$M(t) \stackrel{\text{def}}{=} \exp(e^{sX}) = \exp\left(\mu s + \frac{\sigma^2 t^2}{2}\right),$$

and its cumulant $c_n$ generating function $K(t)$ can be determined as

$$\exp[K(t)] \stackrel{\text{def}}{=} M(t)$$
$$\Rightarrow K(t) = \log M(t)$$

$$= \mu s + \frac{\sigma^2 t^2}{2}.$$

Therefore, values of the cumulant can be calculated as

$$c_n = \begin{cases} \mu & if\ n = 1 \\ \sigma^2 & if\ n = 2 \\ 0 & if\ n \geq 3 \end{cases}.$$

Hyvärinen et al. pointed out that Gaussian distributions were not useful enough to identify the characteristics of variables, because the values of cumulants were equal to 0 where $n \geq 3$.

Furthermore, Hyvärinen et al. (2001) focused on the identifiability of the ICA model. Equation (9) can be written as

$$\mathbf{y} = \mathbf{As}$$

where $\mathbf{A}$ is a $n \times n$ matrix, $\mathbf{y}$ and $\mathbf{s}$ are $n \times 1$ matrices. Comon (1994) and Eriksson and Koivunen (2004) found that $\mathbf{A}$ could be identifiable except for the scale and order of columns. In other words, although $\mathbf{A}$ cannot be estimated perfectly, $\mathbf{A}_{ICA}$ can be estimated and its equation can be represented by

$$\mathbf{A}_{ICA} = \mathbf{ADP} \tag{10}$$

where $\mathbf{D}$ is a diagonal matrix, and $\mathbf{P}$ is a permutation matrix.

Now, in order to estimate $\mathbf{s}$ and $\mathbf{A}$, consider a vector $\mathbf{z}$ as

$$\mathbf{z} = \mathbf{Wy}.$$

$\mathbf{W}$ is called as a demixing matrix, and the estimate of $\mathbf{s}$ is obtained where $\mathbf{W}$ equals to $\mathbf{A}^{-1}$. According to Hyvärinen et al. (2001), the way to obtain the estimates of $\mathbf{W}$ is to hunt for a de-mixing matrix that maximize the independence of each component of $\mathbf{z}$. In other words, the maximization of non-Gaussianity of $z_j$ is the key to estimating the ICA models. As Shimizu (2017) noted, one of the indicators that measures the independence is given by

$$I(\mathbf{z}) = \left\{\sum_{j=1}^{n} H(z_j)\right\} - H(\mathbf{z})$$

where $H(\cdot)$ denotes the entropy of a stochastic variable. The entropy of $\mathbf{z}$ can be written as

$$H(\mathbf{z}) = exp[-log\ p(\mathbf{z})].$$

$I(\mathbf{z})$ is generally termed as mutual information, and its estimator is determined by

$$\hat{I}(\mathbf{z}) = \left[\sum_{j=1}^{n} \frac{1}{r}\sum_{k=1}^{r}\{-log\ p(w_j'\ y_k)\}\right] - \frac{1}{r}\sum_{k=1}^{r}\{-log\ p(\mathbf{W}\ y_k)\},$$

where $w_j'$ is a row vector from the $j$-th row of $\mathbf{W}$. Moreover, $\mathbf{W}$ is identifiable except for the components of the permutation matrix $\mathbf{P}$ and diagonal matrix $\mathbf{D}$. Because $\mathbf{W}$ is an inverse matrix of $\mathbf{A}$, we can obtain $\mathbf{W}_{ICA}$ from equation (10),

$$\mathbf{W}_{ICA} = \mathbf{PDW}. \tag{11}$$

Although the matrices $\mathbf{P}$, $\mathbf{D}$, and $\mathbf{W}$ are all unknown to us, we can estimate the product of these three $\mathbf{W}_{ICA}$ by means of ICA. Moreover, Shimizu, Hoyer, Hyvärinen, and Kerminen (2006) proposed to utilize ICA with a part of graph theory for causal discovery among multiple variables.

*2.2.4 Linear non-Gaussian acyclic model*

Shimizu et al. (2006) demonstrated the way to reveal causal relationships using ICA. They constructed a semi-parametric approach which is termed as "linear non-Gaussian acyclic model (LiNGAM)." The semi-parametric approach is a method which contains characteristics of both parametric and non-parametric approaches. According to Shimizu (2017), it assumes that the functional type is linear (based on the parametric) and does not assume the distribution of independent

variables (based on the non-parametric). It is defined by
$$y_i = \sum_{k(j)<k(i)} b_{ij} y_j + e_i, \tag{12}$$
where $i = 1, \cdots p$, $j = 1, \cdots p$, and $j \neq i$. $y_i$ is a linear combination of its error term $e_i$ and other variables $y_j$. The coefficient $b_{ij}$ represents the existence or scale of the direct causal relationship of $y_j \rightarrow y_i$, and it is known as a connection strength. It is assumed that the error term $e_i$ is given exogenously, its distribution is non-Gaussian, and all the error terms are independent each other.

In equation (12), $k(\cdot)$ denotes a causal order among variables. Shimizu (2014; 2017) defined that the it was an order that no latter variable affected former variables. For instance, when a causality $y_c \rightarrow y_a \rightarrow y_b$ exists, the causal order can be represented as $k(c) = 1$, $k(a) = 2$, and $k(b) = 3$. Equation (12) also can be written in a matrix form as
$$\mathbf{y} = \mathbf{B}\mathbf{y} + \mathbf{e}. \tag{13}$$
It should be noted that $\mathbf{B}$ can be permuted into a lower triangular matrix whose diagonal components are zero. Let us consider an example of LiNGAM of three structural equations,
$$y_1 = -0.08 y_3 + e_1$$
$$y_2 = 0.85 y_1 + e_2$$
$$y_3 = e_3.$$
The set of these equations can be written in the form of equation (13),
$$\begin{bmatrix} y_1 \\ y_2 \\ y_3 \end{bmatrix} = \begin{bmatrix} 0 & 0 & -0.08 \\ 0.85 & 0 & 0 \\ 0 & 0 & 0 \end{bmatrix} \begin{bmatrix} y_1 \\ y_2 \\ y_3 \end{bmatrix} + \begin{bmatrix} e_1 \\ e_2 \\ e_3 \end{bmatrix}$$
$$= \begin{bmatrix} 0 & 0 & 0 \\ -0.08 & 0 & 0 \\ 0 & 0.85 & 0 \end{bmatrix} \begin{bmatrix} y_3 \\ y_1 \\ y_2 \end{bmatrix} + \begin{bmatrix} e_3 \\ e_1 \\ e_2 \end{bmatrix}.$$
Here, we can draw a causal graph of this LiNGAM. For this example, the causal order is determined as $k(3) = 1$, $k(1) = 2$, and $k(2) = 3$, and its causal graph is provided in Figure 4. In the next section, we show the way to identify an accurate LiNGAM using ICA.

*2.2.4 Identification of LiNGAM using ICA*

Shimizu et al. (2006) and Shimizu (2017) demonstrated how to identify $\mathbf{P}$ and $\mathbf{D}$ in equation (11) by means of ICA. For simplicity, consider a LiNGAM of two structural equations,
$$y_1 = e_1$$
$$y_2 = b_{21} y_1 + e_2.$$
The matrix form of this model can be represented as
$$\begin{bmatrix} y_1 \\ y_2 \end{bmatrix} = \begin{bmatrix} 0 & 0 \\ b_{21} & 0 \end{bmatrix} \begin{bmatrix} y_1 \\ y_2 \end{bmatrix} + \begin{bmatrix} e_1 \\ e_2 \end{bmatrix}$$
$$= \left( \begin{bmatrix} 1 & 0 \\ 0 & 1 \end{bmatrix} - \begin{bmatrix} 0 & 0 \\ b_{21} & 0 \end{bmatrix} \right)^{-1} \begin{bmatrix} e_1 \\ e_2 \end{bmatrix}$$
$$= \begin{bmatrix} 1 & 0 \\ -b_{21} & 1 \end{bmatrix}^{-1} \begin{bmatrix} e_1 \\ e_2 \end{bmatrix} \quad (= \mathbf{W}^{-1}\mathbf{e})$$
$$= \begin{bmatrix} 1 & 0 \\ b_{21} & 1 \end{bmatrix} \begin{bmatrix} e_1 \\ e_2 \end{bmatrix} \quad (= \mathbf{A}\mathbf{e}).$$
We now obtain the correct matrix $\mathbf{W}$. Further, by multiplying $\mathbf{W}$ with a diagonal matrix $\mathbf{D}$ whose diagonal components are non-zero, we obtain the product $\mathbf{DW}$ by
$$\mathbf{DW} = \begin{bmatrix} d_{11} & 0 \\ -d_{22} b_{21} & d_{22} \end{bmatrix}.$$

This form is correctly permutated, and it is known that one or more of the diagonal components become zero for the matrices of other forms (Shimizu et al.). Here, we seek such a matrix of $\boldsymbol{W}_{ICA}$ in equation (11) by means of ICA.

Hyvärinen et al. (2010) and Shimizu (2014) introduced the log likelihood of LiNGAM. It is represented by

$$log\ L(\boldsymbol{Y}) = \sum_t \sum_i log\ p_i \left(\frac{y_t - b_i^T y_t}{\sigma_i}\right) - n \sum_i log\ \sigma_i, \qquad (14)$$

where $p_i(\cdot)$ is a probability density function of the standardized error term $e_i/\sigma_i$. Maximizing the likelihood of equation (14) for all possible causal orders seems the most clear-cut way to estimate the matrix $\boldsymbol{B}$. Unfortunately, Hyvärinen et al. did not recommend such an approach because it needed too much data processing. Shimizu et al. (2006) alternatively proposed to apply ICA for that estimation, and we follow this procedure. The first step is the application of ICA. We employ an algorithm which is known as FastICA, proposed by Hyvärinen (1999). Second, the permutation of the estimated matrices is held to make them into being lower triangular. Only these two steps are needed in the estimation of LiNGAM by ICA. Further details of the ICA algorithm for LiNGAM can be found in Hyvärinen et al. (2010) and Shimizu (2014).

*2.2.5 SVAR-LiNGAM approach*

In order to analyze time series data with the LiNGAM approach, Hyvärinen et al. (2010) and Moneta et al. (2013) integrated the approach with an SVAR model. They termed it as a VAR-LiNGAM approach; however, in this paper we call it "SVAR-LiNGAM" after here to hit out on the difference from the conventional VAR. The SVAR-LiNGAM can be represented as

$$\boldsymbol{y}_t = \sum_{h=0}^{p} \boldsymbol{B}_h \boldsymbol{y}_{t-h} + \boldsymbol{u}_t, \qquad (15)$$

where $\boldsymbol{y}_t$ is observed, $\boldsymbol{u}_t$ is an error term which is given exogenously. $\boldsymbol{B}_h$ denotes matrices of the connection strength coefficients with a time lag $h$. As $h$ can be zero, $\boldsymbol{B}_h$ includes both instantaneous and lagged effects. As with equation (13), $\boldsymbol{B}_h$ can be permuted into a lower triangular matrix whose diagonal components are 0. This is because LiNGAM assumes a directed acyclic graph (DAG) structure.

General econometric methods, such as the conventional VAR and VECM approach, assume that its error terms follow a Gaussian distribution. Although such an assumption is not satisfied in many cases, it is often considered to be robust, or ignorable. This can be considered as a problem of the VAR and VECM. Especially for crypto-asset data, it is difficult to construct a multivariate time-series model, since the price often volatiles and surges. Here, the SVAR-LiNGAM approach appears to be a better way to capture the appropriate estimates, as the model assumes non-Gaussianity and utilizes its information.

**3. Empirical results Data and methodology**

*3.1 Results of unit root and cointegration tests*

*3.1.1 Unit root test*

The time-series data of the four exchange rates were examined to determine whether each process has a unit root. First, we tested the original time series by the ADF test. Appropriate lag lengths for the four processes were selected by Schwarz Information Criterion (Schwarz, 1978).

Table 2 shows the results of the ADF tests for the four series. According to the results obtained, $H_0$ could not be rejected in terms of all the test statistics. It was suggested that all four series had a unit root. Second, the ADF tests were also held on the first-difference series. Table 3 shows the results for the four first-difference series. According to the hypothesis testing, $H_1$ was rejected in terms of all the test statistics on the 1 % level of significance. It was suggested that all four first-difference series did not have a unit root. These two results of hypothesis testing indicated that all the original series, $BTC_t$, $ETH_t$, $XRP_t$, and $SPT_t$, each had a unit root.

*3.1.2 Cointegration test*

Second, we calculated the four eigenvalues $\hat{\lambda}_1$, $\hat{\lambda}_2$, $\hat{\lambda}_3$, and $\hat{\lambda}_4$. Using the selected lag lengths, we examined the time series by the two cointegration tests. We checked the four null hypotheses for the time series and obtained test statistics: $\hat{\lambda}_{trace}$ and $\hat{\lambda}_{eigen}$. Table 3 shows the results of the hypothesis testing of cointegration for the four-variable model.

The hypotheses $r = 0$, $r \leq 1$, and $r \leq 2$ were rejected at the 1 % significance level, and the hypothesis $r \leq 3$ could not be rejected in the two tests. This means that the four variables appear to have three cointegrating relationships.

*3.2 Results of the SVAR-LiNGM and its network*

*3.2.1 Results of Gaussianity and serial-correlation tests*

We first examined the non-Gaussianity of the error terms of the reduced form of VECM. Figure 4 shows histograms and Q-Q plots of the error terms. We conducted several Gaussianity tests, such as Shapiro-Wilk, Shapiro-Francia, and Jarque-Bera tests. The results indicated that all four error terms were non-Gaussian at the 1 % significance level (see Table 14). Here, the four processes were applicable for the SVAR-LiNGAM analysis because they met the necessary and sufficient condition: non-Gaussianity of the error terms. In addition to the original application of SVAR-LiNGAM by Moneta et al. (2013), we added serial correlation tests for the residuals. Non-stationary time series

*3.2.2 Estimated instantaneous causal effect matrix and lagged causal effect matrix*

We applied the SVAR-LiNGAM analysis to estimate the causal relationships among the four time series variables. We estimated the instantaneous and lagged effects by conducting 1,000 iterations of bootstrapping.

The lag for the four-variable VAR model was selected as three by SIC. Based on equations (2) and (15), the four-variable VAR (3) and SVAR-LiNGAM (3) can be written by

$$\begin{pmatrix} SPT_t \\ BTC_t \\ ETH_t \\ XRP_t \end{pmatrix} = \boldsymbol{\gamma} + \boldsymbol{\Gamma}_1 \begin{pmatrix} SPT_{t-1} \\ BTC_{t-1} \\ ETH_{t-1} \\ XRP_{t-1} \end{pmatrix} + \boldsymbol{\Gamma}_2 \begin{pmatrix} SPT_{t-2} \\ BTC_{t-2} \\ ETH_{t-2} \\ XRP_{t-2} \end{pmatrix} + \boldsymbol{\Gamma}_3 \begin{pmatrix} SPT_{t-3} \\ BTC_{t-3} \\ ETH_{t-3} \\ XRP_{t-3} \end{pmatrix} + \boldsymbol{\varepsilon}_t, \qquad (16)$$

$$\boldsymbol{B}_0 \begin{pmatrix} SPT_t \\ BTC_t \\ ETH_t \\ XRP_t \end{pmatrix} = \boldsymbol{c} + \boldsymbol{B}_1 \begin{pmatrix} SPT_{t-1} \\ BTC_{t-1} \\ ETH_{t-1} \\ XRP_{t-1} \end{pmatrix} + \boldsymbol{B}_2 \begin{pmatrix} SPT_{t-2} \\ BTC_{t-2} \\ ETH_{t-2} \\ XRP_{t-2} \end{pmatrix} + \boldsymbol{B}_3 \begin{pmatrix} SPT_{t-3} \\ BTC_{t-3} \\ ETH_{t-3} \\ XRP_{t-3} \end{pmatrix} + \boldsymbol{u}_t, \qquad (17)$$

respectively. Table 5 shows the estimates of the lagged effects $\hat{\mathbf{\Gamma}}$ from the conventional VAR (3) model, and Table 6 shows those of the instantaneous and lagged effects from the SVAR-LiNGAM. Figure 4 illustrates the causal graph of equation (17). As the instantaneous effects $\mathbf{B}_0$ could be considered as the causal structure, the results indicated that the order was $SPT_t \rightarrow BTC_t \rightarrow ETH_t \rightarrow XRP_t$. This is because an SVAR-LiNGAM with no lagged effects is identical with the original LiNGAM. In terms of the lagged effects from $t-1$, $SPT_{t-1}$ had a strong and positive impact on $SPT_t$, and negative impacts on $BTC_t$ and $XRP_t$. It should be noted that the direct causal relationship between $SPT_{t-1}$ and $ETH_t$ was much weaker than that among other variables. In addition, $BTC_{t-1}$ seemed to affect $SPT_t$ positively, matching indications in the former VECM analysis. For the period $t-2$, $SPT_{t-2}$, $BTC_{t-2}$, and $XRP_{t-2}$ impacted on the $XRP_t$ positively. $BTC_{t-2}$ also gave a positive impact on $BTC_t$, whereas it affected $ETH_t$ negatively. $ETH_{t-2}$ positively affected its current value $ETH_t$, on the other hand, it had a negative impact on $XRP_t$. $XRP_{t-2}$ only affected its current price $XRP_t$. For the lagged effect at $t-3$, $BTC_{t-3}$ gave a negative effect on $ETH_t$, whereas both $ETH_{t-3}$ and $XRP_{t-3}$ only affected their current respective values. These results implied that the $EUR/USD$ spot rate at $t-1$ affected prices of the crypto assets negatively, although the spot rate at $t$ affected them positively in accordance with the causal order: the spot rate, Bitcoin, Ethereum, and Ripple.

Comparing the estimates of the SVAR-LiNGAM with those of the VAR model of equation (16), we found differences between their estimated coefficients. For example, the coefficients $\hat{\mathbf{\Gamma}}_1$ of $SPT_{t-1} \rightarrow BTC_t, ETH_t, XRP_t$ and $BTC_{t-1} \rightarrow ETH_t, XRP_t$ in the VAR model were positive; however, those of $\hat{\mathbf{B}}_1$ in the SVAR-LiNGAM were negative. Such difference was caused by a systematic dilemma of the conventional VAR model, which regarded the instantaneous effect as lagged. The SVAR approach revealed that the lagged effects of $SPT_{t-1}$ were negative, although they were thought to be positive in the VAR. In addition, the coefficients $\hat{\mathbf{\Gamma}}_1$ of $SPT_{t-1} \rightarrow ETH_t$ and $ETH_{t-1} \rightarrow XRP_t$ seemed to be spurious, because their estimated coefficients in the SVAR were much smaller and close to 0. As Hyvärinen et al. (2010) noted, it was suggested that the conventional VAR model was not a sufficient approach to analyze causal relationships among financial variables, especially crypto-asset data.

*3.2.3 Impulse response function of the SVAR-LiNGAM*

Figure 5 shows the results of the IRF of the SVAR-LiNGAM (3) with 99-percent confidence bands. We could obtain the results which had similar tendencies with those of the VECM (see Figure 5). Subsequently, we split the whole period into two (2016/12/04 - 2018/01/31 and 2018/02/01 - 2019/10/31) to allow for the possibility of a trend break. Figure 9 shows the results of the IRF of the SVAR-LiNGAM (3) in comparison with the whole period and two subperiods. As noted in the section 3.1.4, $SPT_t$ affected all four variables immediately. Additionally, the results obtained from the structural model indicated that $SPT_t$ affected them less in the latter period. For the $BTC_t$ shock, $SPT_t$, $ETH_t$ and $XRP_t$ were affected in a day, and the effect remained permanently. Although the effect of $BTC_t$ to $SPT_t$ was limited, its scale slightly increased in the second subperiod. For the $ETH_t$ shock, $ETH_t$ and $XRP_t$ were affected slightly, and the effect diminished in four days. For the $XRP_t$ shock, only $XRP_t$ was affected and the effect diminished in four days. These results imply that the spot rate has been the most influential, but the scale of its effect has been falling during these two years. On the other hand, Bitcoin has been getting more influential and enough to affect the $EUR/USD$ spot rate.

## 4. Conclusions

Apart from the conventional VAR or VECM approach, we also applied the SVAR-LiNGAM to illustrate the causal relationships among the four variables. It was notable that the causal order, $SPT_t \rightarrow BTC_t \rightarrow ETH_t \rightarrow XRP_t$, was obtained by this approach. All the instantaneous effects were strongly positive; however, the three lagged effects from $t-1$ were negative. Here, we found a clear difference in the results obtained in the conventional VECM and the SVAR-LiNGAM. As the former approach did not take the instantaneous effects into account in its system, it could not capture the appropriate causal graph among the variables. It was remarkable that spurious and wrongly lagged effects existed in the conventional VAR, and such problems could be removed by the application of the SVAR-LiNGAM.

Moreover, it was notable that $BTC_{t-1}$ positively affected $SPT_t$. This implies that Bitcoin can influence the $EUR/USD$ spot rate positively with a one-day time lag. It was also remarkable that the instantaneous effects were all positive, whereas the spot rate at $t-1$ affected prices of the three crypto assets negatively. Here, further researches and analyses should be followed to clarify why and how such causal relationships exist in the crypto-asset economy.

## Acknowledgment

T.K. wish to express his gratitude to the late Dr. Lukáš PICHL for helpful suggestions and discussions. This work was supported by JSPS KAKENHI Grant Number 17K01270, 20K01752, NOMURA Foundation.

**Figures**

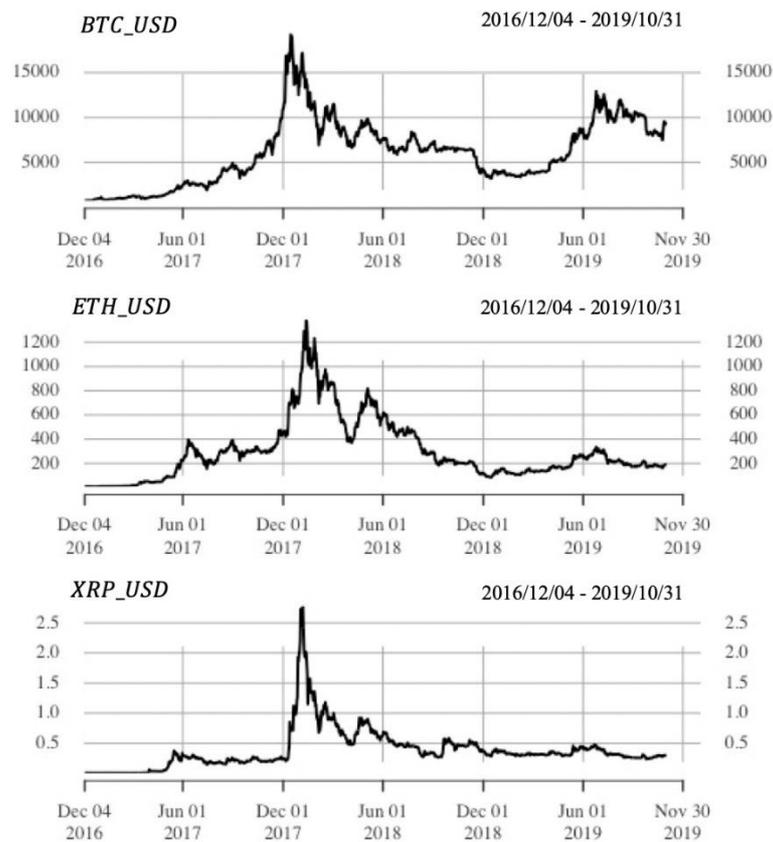

*Fig. 1.* Historical price data of Bitcoin, Ethereum, and Ripple in USD.

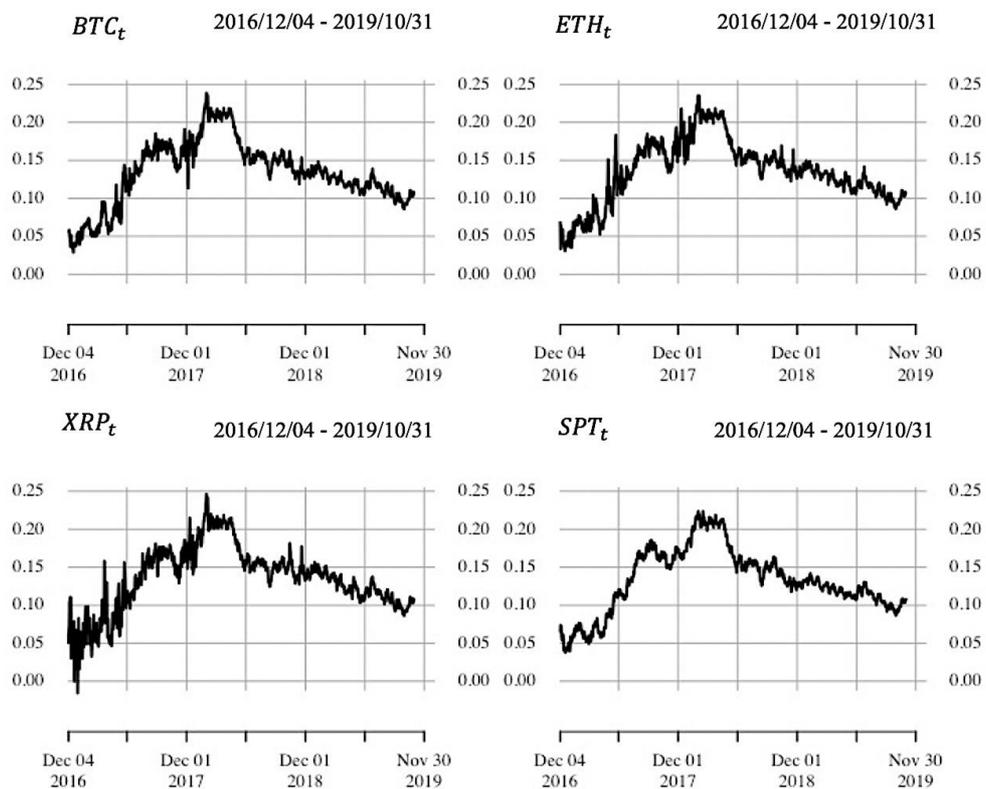

*Fig. 2.* Historical price data of the three crypto-asset exchange rates and the FX spot rate of EUR/USD.

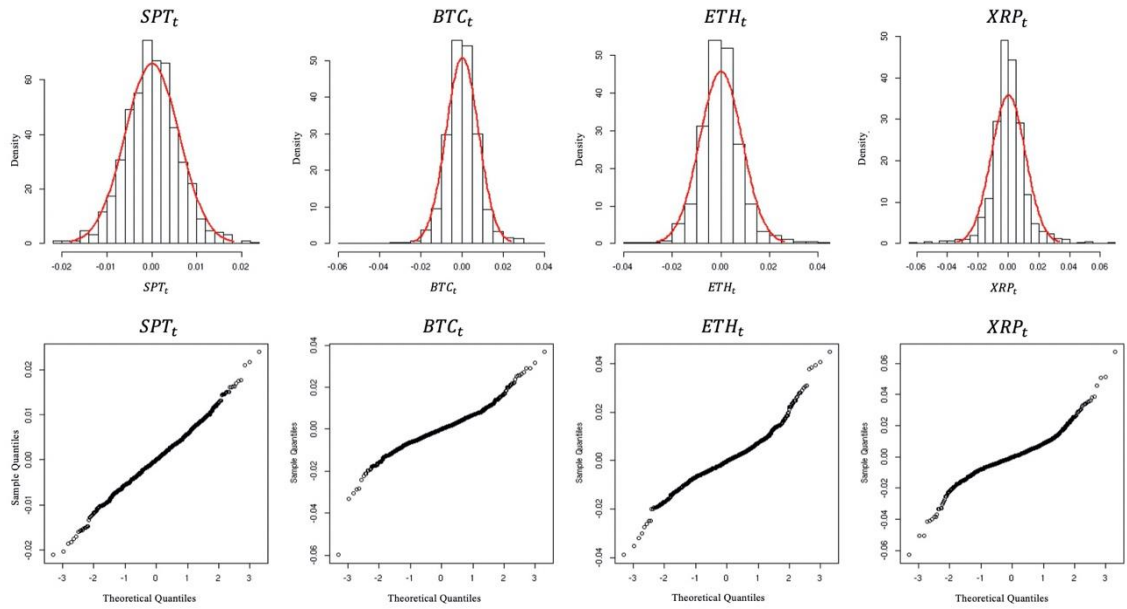

*Fig. 3.* Histograms and plots of error terms of the four-variable VECM (3).

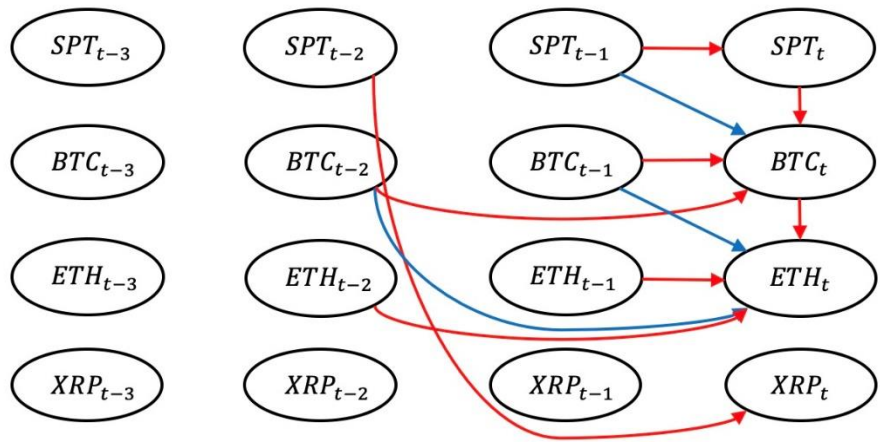

*Fig. 4.* Graphical representation of the four variables of the SVAR-LiNGAM

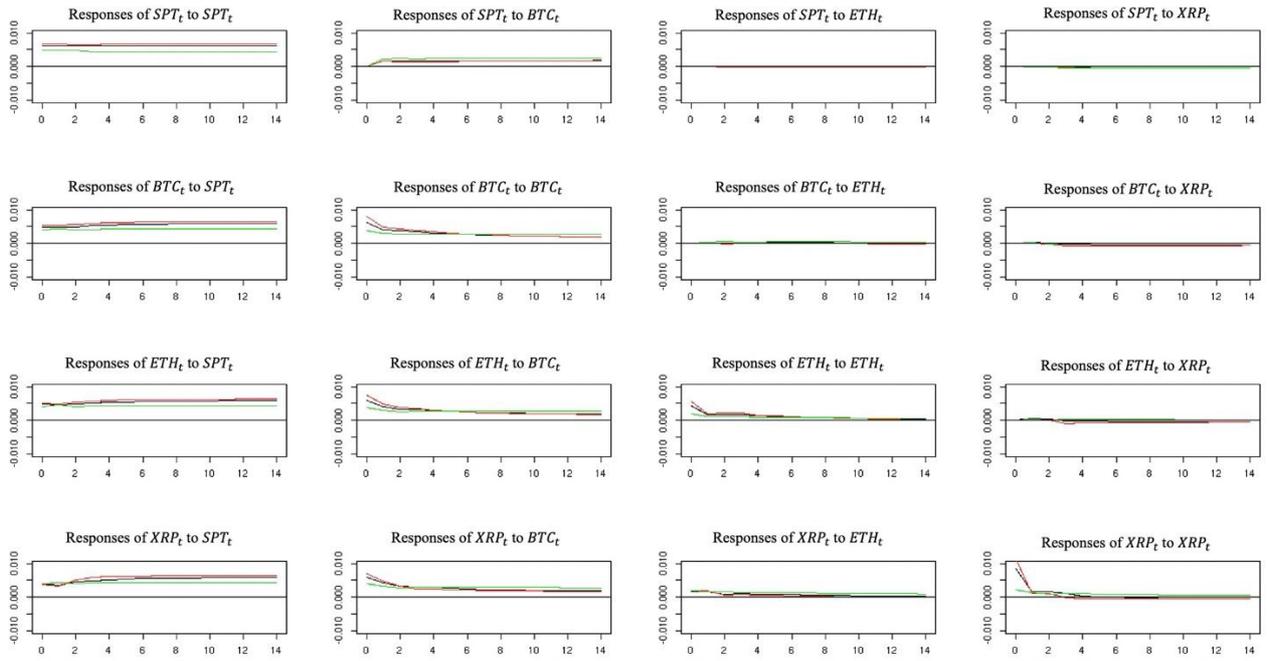

*Fig. 5*. Plots of the impulse response function of the SVAR-LiNGAM.

**Tables**

Table 1
Some descriptive statistics.

| Variables | N | Min | Q1 | Median | Mean | Q3 | Max | S.D. | Skewness | Kurtosis |
|---|---|---|---|---|---|---|---|---|---|---|
| Original $SPT_t$ | 735 | 0.04 | 0.11 | 0.13 | 0.13 | 0.16 | 0.22 | 0.04 | -0.02 | -0.42 |
| Weekend-filled $SPT_t$ | 1062 | 0.04 | 0.11 | 0.13 | 0.13 | 0.16 | 0.22 | 0.04 | -0.04 | -0.45 |
| $BTC_t$ | 1062 | 0.03 | 0.11 | 0.13 | 0.13 | 0.16 | 0.24 | 0.04 | -0.15 | -0.23 |
| $ETH_t$ | 1062 | 0.03 | 0.11 | 0.14 | 0.14 | 0.16 | 0.24 | 0.04 | -0.15 | -0.22 |
| $XRP_t$ | 1062 | -0.02 | 0.11 | 0.14 | 0.14 | 0.16 | 0.25 | 0.04 | -0.26 | 0.00 |

*Note: Original $SPT_t$ is the original data without filling the weekend missing values. Weekend-filled $SPT_t$ is the data whose missing values are complemented by the closing price on Fridays. The original and complemented data of $SPT_t$ show close numbers in summary statistics, and it is indicative of unbiasedness by this complementation. The following tests including $SPT_t$ are analyzed by Weekend-filled $SPT_t$.*

Table 2
Results of the ADF tests on the original and first-difference series.

| Variables | Lag | Test statistics | Variables | Lag | Test statistics |
|---|---|---|---|---|---|
| $SPT_t$ | 1 | -1.914 | $\Delta SPT_t$ | 1 | -24.313*** |
| $BTC_t$ | 2 | -2.250 | $\Delta BTC_t$ | 1 | -28.164*** |
| $ETH_t$ | 2 | -2.403 | $\Delta ETH_t$ | 1 | -30.350*** |
| $XRP_t$ | 3 | -2.132 | $\Delta XRP_t$ | 2 | -27.670*** |

*Note: \*\*\* denotes significance at 1% level. We considered the constant model specification because the estimates of a trend term were small and insignificant when we examined the four processes as the trend model in the ADF tests.*

Table 3
Results of the Johansen cointegrating test for ($\Delta SPT_t, \Delta BTC_t, \Delta ETH_t, \Delta XRP_t$).

| $y_t$ | Lag | Eigenvalues $\hat{\lambda}_1, \hat{\lambda}_2, \hat{\lambda}_3, \hat{\lambda}_4$ | Ranks | Statistics $\hat{\lambda}_{trace}$ | $\hat{\lambda}_{eigen}$ |
|---|---|---|---|---|---|
| ($\Delta SPT_t, \Delta BTC_t,$ $\Delta ETH_t, \Delta XRP_t$) | 3 | 0.162 | r=0 | 367.90*** | 186.50*** |
| | | 0.087 | r≤1 | 181.41*** | 96.70*** |
| | | 0.074 | r≤2 | 84.71*** | 81.61*** |
| | | 0.003 | r≤3 | 3.10 | 3.10 |

*Note: \*\*\* denotes significance at 1% level.*

Table 4
Results of normality tests.

| Variables | Kurtosis | p-values | | |
|---|---|---|---|---|
| | | Shapiro-Wilk | Shapiro-Francia | Jarque-Bera |
| $SPT_t$ | 0.640 | 0.004 | 0.003 | 0.000 |
| $BTC_t$ | 4.893 | 0.000 | 0.000 | 0.000 |
| $ETH_t$ | 3.273 | 0.000 | 0.000 | 0.000 |
| $XRP_t$ | 4.920 | 0.000 | 0.000 | 0.000 |

Table 5
Coefficient matrices $\hat{\Gamma}$ of lagged effects from the conventional VAR (3) model.

| Variables | $\hat{\Gamma}_1$ | | | |
|---|---|---|---|---|
| | $SPT_{t-1}$ | $BTC_{t-1}$ | $ETH_{t-1}$ | $XRP_{t-1}$ |
| $SPT_t$ | 0.8086*** | 0.2496* | 0.0182 | 0.0015 |
| $BTC_t$ | 0.2823** | 0.5888*** | 0.0430 | 0.0242 |
| $ETH_t$ | 0.2261 | 0.2561* | 0.3652*** | 0.0590 |
| $XRP_t$ | 0.0381 | 0.1873 | 0.3244** | 0.2016 |
| | $\hat{\Gamma}_2$ | | | |
| | $SPT_{t-2}$ | $BTC_{t-2}$ | $ETH_{t-2}$ | $XRP_{t-2}$ |
| $SPT_t$ | 0.0610 | -0.0860 | -0.0312 | -0.0017 |
| $BTC_t$ | -0.0492 | 0.1508 | -0.0577 | -0.0150 |
| $ETH_t$ | 0.0403 | -0.1682* | 0.2200** | -0.0141 |
| $XRP_t$ | -0.1979** | 0.0072 | -0.0887 | 0.1369 |
| | $\hat{\Gamma}_3$ | | | |
| | $SPT_{t-3}$ | $BTC_{t-3}$ | $ETH_{t-3}$ | $XRP_{t-3}$ |
| $SPT_t$ | 0.0539 | -0.0582 | 0.0147 | -0.0306 |
| $BTC_t$ | -0.0217 | 0.0264 | 0.0773 | -0.0489 |
| $ETH_t$ | -0.0469 | -0.0841 | 0.2067** | -0.0686 |
| $XRP_t$ | -0.0130 | -0.1031 | 0.0484 | 0.0568 |

*Note: *, **, and *** denote significance at 10%, 5%, 1% level, respectively.*

Table 6

Coefficient matrices $\hat{B}$ of instantaneous and lagged effects from the SVAR-LiNGAM (3).

| Variables | $\hat{B}_0$ | | | |
|---|---|---|---|---|
| | $SPT_t$ | $BTC_t$ | $ETH_t$ | $XRP_t$ |
| $SPT_t$ | 0.0000 | 0.0000 | 0.0000 | 0.0000 |
| $BTC_t$ | 0.7893** | 0.0000 | 0.0000 | 0.0000 |
| $ETH_t$ | 0.0275 | 0.9531*** | 0.0000 | 0.0000 |
| $XRP_t$ | -0.0731 | 0.5303 | 0.4100 | 0.0000 |
| | $\hat{B}_1$ | | | |
| | $SPT_{t-1}$ | $BTC_{t-1}$ | $ETH_{t-1}$ | $XRP_{t-1}$ |
| $SPT_t$ | 0.8086*** | 0.2496 | 0.0182 | 0.0015 |
| $BTC_t$ | -0.3559** | 0.3918** | 0.0286 | 0.0231 |
| $ETH_t$ | -0.0652 | -0.3120** | 0.3238** | 0.0359 |
| $XRP_t$ | -0.1452 | -0.2117 | 0.1532 | 0.1647 |
| | $\hat{B}_2$ | | | |
| | $SPT_{t-2}$ | $BTC_{t-2}$ | $ETH_{t-2}$ | $XRP_{t-2}$ |
| $SPT_t$ | 0.0610 | -0.0860 | -0.0312 | -0.0017 |
| $BTC_t$ | -0.0974 | 0.2187** | -0.0331 | -0.0137 |
| $ETH_t$ | 0.0856 | -0.3096*** | 0.2759** | 0.0003 |
| $XRP_t$ | 0.2120** | -0.0101 | -0.1506 | 0.1505 |
| | $\hat{B}_3$ | | | |
| | $SPT_{t-3}$ | $BTC_{t-3}$ | $ETH_{t-3}$ | $XRP_{t-3}$ |
| $SPT_t$ | 0.0539 | -0.0582 | 0.0147 | -0.0306 |
| $BTC_t$ | -0.0642 | 0.0723 | 0.0657 | -0.0247 |
| $ETH_t$ | -0.0277 | -0.1076 | 0.1327 | -0.0211 |
| $XRP_t$ | 0.0217 | -0.0869 | -0.0763 | 0.1086 |

*Note: ** and *** denote significance at 5% and 1% level, respectively.*